\documentstyle[12pt]{ioplppt}
\begin{document}
\jl{6}
\title{New first integral for twisting type-N vacuum gravitational
 fields with two non-commuting Killing vectors}[New first integral for 
twisting type-N fields] 

\author{F J Chinea}
\address{Departamento de F\'{\ii}sica Te\'orica II, Facultad de Ciencias
F\'{\ii}sicas, Universidad Complutense, 28040 Madrid, Spain}

\begin{abstract}
A new first integral for the equations corresponding to twisting type-N
vacuum gravitational fields with two non-commuting Killing vectors is
introduced. A new reduction of the problem to a complex second-order
ordinary differential equation is given. Alternatively, the mentioned
first integral can be used in order to provide a first integral of
the second-order complex equation introduced in a previous treatment
of the problem.  
\end{abstract}

\section{Introduction\label{intro}}
It is known from the peeling-off property that 
the dominant term in an asymptotic expansion of the gravitational field
is of algebraic type N. Thus, such fields can be considered as
containing the main features of gravitational radiation far from the
sources. However, twist-free type-N fields are known to contain
unavoidable singularities. The only explicitly known type-N vacuum
field with twist \cite{H1} suffers from similar problems. In the
quest for considering a purely type-N field throughout the whole
spacetime as a theoretical model for realistic ones produced by
compact sources, there remains the hope that twisting fields with the
appropriate behaviour may eventually be found. In spite of numerous
efforts, no new explicit fields are known. The information gathered
on twisting type-N fields, especially when additional symmetries are
considered, can be seen in references such as \cite{DKS}-\cite{FPP2}.

The maximum number of independent isometries for such
gravitational fields can be shown to be two, in which case the
corresponding Lie algebra is necessarily non-Abelian \cite{C}.
Although the approach used in \cite{H1} was based on algebraic
properties of some exterior form equations, the solution found by
Hauser turns out to possess one Killing vector and one
{\it homothetic} Killing vector. This naturally led to a more detailed
investigation of symmetries of such a type. The special case of two
proper Killing vectors was treated in \cite{PRD88}, where a
reduction of the problem to a second-order complex
differential equation was shown. By further manipulation, it was
also shown there that the problem can be reduced to the study of a
single third-order real equation. In what follows, we first discuss in
the next section the main features and results of the previous
treatment. Within this framework, we then introduce in the final
section a new first integral of the field equations; it gives
rise to a more direct reduction in terms of the original
variables appearing in the metric, while it can also be used as
a (previously unnoticed) first integral of the second-order
complex equation considered in \cite{PRD88}. Our aim in
pursuing the present approach is twofold: firstly, to cast the
problem in a form more amenable to numerical treatment, and
secondly, to open the way for the intriguing possibility of
final reduction to a second-order {\it real} equation.

\section{Twisting type-N vacuum fields with two non-commuting Killing 
vectors}

The reduction of the field and symmetry equations described in this
section was carried out in \cite{PRD88} by means of a compact version
of the Cartan structure equations \cite{PRL84}, in terms of an 
$sl(2,C)$-valued 1-form connection; the empty space Einstein equations
then take a particularly simple form. The conditions for the existence
of two non-commuting Killing vectors and the final expression for such
Killing fields were performed at the tetrad rather than at the metric
level, by using techniques modelled on similar ones in gauge
theories \cite{CQG88}. Based on \cite{C}, it was required that the Lie
algebra of isometries took the form
\begin{equation}
[\xi _{1},\xi _{2}] =\xi _{1}. \label{Lie}
\end{equation}
The connection 1-form was chosen to be in a gauge of maximum simplicity,
following the gauge used in \cite{H1}. Such a gauge is compatible with
the requirement that the field be of type N. It leads to the vanishing
of all components of the Weyl spinor, except for $\Psi _{4}.$ The
metric is expressed in terms of the null tetrad $\{l,n,m,\bar{m}\}$
(where $l$ and $n$ are real 1-forms, while $m$ is a complex one, with
complex conjugation being denoted by a bar) by means of the usual
relation 
\begin{equation} 
g=-2l\otimes _{s}n+2m\otimes _{s}\bar{m}, \label{metric}
\end{equation}
where $\otimes _{s}$ means the symmetrized tensor product. The 1-form
$l$ is the repeated principal null eigenform, and its twist is
assumed not to vanish. The four independent spacetime
coordinates are $\{u,\sigma,\zeta,\bar{\zeta}\},$ where $u$ and
$\sigma$ are real, while $\zeta$ is complex. Details of the procedure
need not concern us here; they are given in \cite{CQG88} and
\cite{PRD88}. The Killing vectors can be shown to be
\begin{equation} 
\xi _{1}= i\partial _{\zeta} -i\partial _{\bar{\zeta}} \label{xi1}
\end{equation}
and
\begin{equation} 
\xi _{2}= \zeta \partial _{\zeta} +\bar{\zeta}\partial _{\bar{\zeta}}.
 \label{xi2}
\end{equation}
The tetrad can be expressed as \footnote{
The tetrad was denoted by ${k,m,t,\bar{t}}$ in \cite{PRD88};
the correspondence with the notation used here is the following:
$k\rightarrow -l,$ $m\rightarrow n,$ $t\rightarrow m.$
} 
\begin{equation}
l=(\zeta + \bar{\zeta})\d u +D \d \zeta+ \bar{D} \d \bar{\zeta}\label{l}
\end{equation}
\begin{equation}
\fl n=(\zeta + \bar{\zeta})^{-1}\d \sigma +(-\sigma +F\bar{D}-M)
(\zeta + \bar{\zeta})^{-2}\d \zeta + (-\sigma +\bar{F}D-\bar{M})
(\zeta + \bar{\zeta})^{-2}\d \bar{\zeta}
\label{n} 
\end{equation}
\begin{equation}
m=\bar{D}^{-1}\bar{M}\d u -\sigma (\zeta + \bar{\zeta})^{-1}\d \zeta+
\bar{M}(\zeta + \bar{\zeta})^{-1}\d \bar{\zeta},
\label{m}
\end{equation}
where $D,$ $M$ and $F$ are complex functions of the real independent
variable $u$ only. Due to the first Bianchi identities, the
requirement that the resulting metric be of type N, the conditions
for the existence of the Killing fields, and the vacuum field
equations, these functions are subject to the following system of
equations: 
\begin{equation} M_{u}=-F, \label{Mu}  \end{equation}
\begin{equation}
F_{u}=-2F\bar{D}^{-1}, \label{Fu}  
\end{equation}
\begin{equation}
D_{u}=1-D^{-1}M, \label{Du} 
\end{equation}
\begin{equation}
2M-2\bar{M}-2F\bar{D}+2\bar{F}D+F\bar{M}-\bar{F}M=0,
\label{constraint} 
\end{equation}
where the subscript denotes a derivative with respect to $u.$
Note that \eref{constraint} takes the form of a first integral of the
previous equations \eref{Mu}-\eref{Du} (in the sense that the
derivative of its left-hand side with respect to $u$ vanishes when
those first three equations are taken into account). Two points should
be stressed, however; firstly, \eref{constraint} appears by itself as one
of the combined field and symmetry equations, rather than being
searched for; and secondly, its right-hand side is not an arbitrary
constant, but has to take the specific value zero.

If we define the new independent variable $s=\ln v$, with
\begin{equation}
v(u)=\int F(u)\bar{F(u)}\d u,
\label{v} 
\end{equation}
and the new complex dependent variable $z$ as
\begin{equation}
z=2v^{-1}M,
\label{z} 
\end{equation}
then the set of equations \eref{Mu}-\eref{constraint} reduce to the
single second-order autonomous ordinary differential equation
\begin{equation}
\ddot{z}+\dot{z}+\frac{2(z+\dot{z})^{2}(\bar{z}+\dot{\bar{z}})}
{z\bar{z}+\bar{z}\dot{z}+\dot{z}}=0,
\label{edoz} 
\end{equation}
where dots denote derivatives with respect to the real independent
variable $s.$ By rewriting the set \eref{Mu}-\eref{constraint} as
equations in the independent variable $v,$ it can be easily seen that
$M,$ $D$ and $F$ are explicitly given algebraically in terms of
$z,$ $\bar{z}$ and their first derivatives (the second and successive
derivatives of $z$ and $\bar{z}$ can be expressed in terms of the
first-order ones by means of \eref{edoz} and its complex conjugate).
Finally, the original independent variable $u$ is simply given by a
quadrature, corresponding to the inversion of \eref{v}. The details
are given in \cite{PRD88}. 

By further manipulation of \eref{edoz} (involving a Legendre
transformation and the compatibility conditions for an associated linear
system), it was shown in \cite{PRD88} that the problem can be
reduced to a single, third-order, {\it real} ordinary differential
equation. Similar techniques have been subsequently used in \cite{Herlt}
and \cite{LY} in the more general case of the existence of one
Killing vector and one {\it homothetic} Killing vector, showing
also a reduction to a final, third-order equation. It is rather
remarkable that the simplest such equation was found in
\cite{Herlt} for one specific case of the existence of a
{\it proper} homothetic Killing vector (i.e. a case not reducible
to the situation considered in the present paper). The reduction
to third-order equations and the relations among different approaches
have been further investigated in \cite{FPP1}, \cite{LE}, and
\cite{EL}. 

For completeness, the remainder of this section will be used for
recording some important quantities that appear in the present problem.
They can be computed in a straightforward fashion from the tetrad
\eref{l}-\eref{m} (taking into account \eref{Mu}-\eref{constraint}
when derivatives are involved), especially if one uses some computer
algebra software. The only non-vanishing spin coefficients are the
following:
\begin{equation}
\lambda = \frac{F\bar{D}}{(\sigma \bar{D}+\bar{M}D)(\zeta +
\bar{\zeta})} \label{lambda}
\end{equation}
\begin{equation}
\nu =  \frac{F\bar{M}}{(\sigma \bar{D}+\bar{M}D)(\zeta +
\bar{\zeta})^2} \label{nu}
\end{equation}
\begin{equation}
\rho = -\frac{\bar{D}(\zeta +
\bar{\zeta})}{(\sigma \bar{D}+\bar{M}D)} \label{rho}
\end{equation}
\begin{equation}
\tau = -\frac{\bar{M}}{(\sigma \bar{D}+\bar{M}D)}. \label{tau}
\end{equation}
As $\rho = \theta + i\omega,$ where $\theta$ is the expansion and
$\omega$ is the twist of the congruence defined by $l,$ the twist is given
by the following expression:
\begin{equation}
\omega = \frac{i}{2}(\zeta +
\bar{\zeta})\frac{M\bar{D}^2-\bar{M}D^2}{(\sigma \bar{D}+\bar{M}D)
(\sigma D+M\bar{D})}.
\label{twist} 
\end{equation}
All components of the Weyl spinor vanish, except for $\Psi _{4},$
as corresponds to the type-N character in the gauge we are using. The
latter is given by
\begin{equation}
\Psi _{4} =  \frac{2F}{(\sigma \bar{D}+\bar{M}D)(\zeta +
\bar{\zeta})^2}. \label{psi}
\end{equation}
From \eref{twist} and \eref{psi}, we should keep in mind for
non-triviality that 
\begin{equation}
{\rm non}{\rm -}{\rm flat}\,{\rm spacetime} \Longleftrightarrow F \neq 0,
\label{truepsi}
\end{equation}
while
\begin{equation}
{\rm non}{\rm -}{\rm vanishing}\,{\rm twist} \Longleftrightarrow
M\bar{D}^2-\bar{M}D^2 \neq 0. \label{truetwist}
\end{equation}
Note that the condition \eref{truepsi} is precisely what is required
for the coordinate change \eref{v} to be defined.

As the twist measures the failure of the 1-form $l$ to be
Frobenius integrable, it is sometimes convenient to consider the
following alternative expression to \eref{twist}:
\begin{equation}
\d l\wedge l=-\frac{1}{D\bar{D}}(M\bar{D}^2-\bar{M}D^2)\d u \wedge \d
\zeta \wedge \d \bar{\zeta}. \label{Frobenius} 
\end{equation}

\section{New first integral and reduction}
The present author realized recently that the system of equations
\eref{Mu}-\eref{Du} admit the following first integral, which
went unnoticed in \cite{PRD88}:
\begin{equation}
F\bar{D}^2+\bar{F}D^2-2M\bar{M}=k, 
\label{new}
\end{equation} 
where $k$ is an arbitrary real constant. As an alternative to the
reduction to a final second-order complex equation given in
\cite{PRD88} (equation \eref{edoz} above), in the manner discussed in the
previous section, the following new possibility presents itself: from
\eref{Du} one can write
\begin{equation}
M=D-DD_{u},
\label{M}
\end{equation}
and then, upon differentiation of \eref{M} with respect to $u$ and 
substitution into \eref{Mu}, $F$ can be expressed as
\begin{equation}
F=DD_{uu}+D_{u}^2-D_{u}.
\label{F}
\end{equation}
By substituting $M,$ $\bar{M},$ $F$ and $\bar{F}$ as given by
\eref{M} and \eref{F} (and their complex conjugates) into the
constraint equation \eref{constraint} and the first integral
\eref{new}, one ends up with two real equations where only $D,$
$\bar{D},$ and their derivatives up to the second order appear.
They can be solved for $D_{uu}$ (and its complex conjugate). The
result is the following:
\begin{equation}
D_{uu}=\frac{A}{B},
\label{Duu}
\end{equation}
where
\begin{eqnarray}
 A=-\bar{D}^2D_{u}^3+D\bar{D}\bar{D}_{u}D_{u}^2-3D\bar{D}D_{u}^2+
D\bar{D}D_{u}\bar{D}_{u}  \nonumber\\
-2D^2 D_{u}+\bar{D}^2
D_{u}+D\bar{D}D_{u}+kD_{u}+2D^2+k, \label{A}
\end{eqnarray}
and
\begin{equation}
B=D\bar{D}(D\bar{D}_{u}+\bar{D}D_{u}+D+\bar{D}). 
\label{B}
\end{equation}
Finally, there remains equation \eref{Fu}, which is third order in $D$.
But third-order derivatives can be obtained from \eref{Duu}. When this
is done, \eref{Fu} is satisfied identically. In other words, the only
equation to be solved is precisely the second-order complex
equation \eref{Duu}. While this equation is rather complicated, it has
the following property: it determines $D$ as a function of the original
independent variable $u,$ without the need of transformations such as
\eref{v}. It seems that this will allow for a more direct numerical
treatment of the twisting type-N problem, as compared with an approach
based on \eref{edoz} or its associated third-order real equation. On
the other hand, the new integral \eref{new} can also be used in
connection with \eref{edoz}: By carrying out the changes discussed in
the previous section, one ends up with additional information about
\eref{edoz}. Namely, it admits the following first integral:
\begin{equation}
I_{0}=\frac{\e ^{2s}}{\bar{z}+\dot{\bar{z}}}
(z\bar{z}+z\dot{\bar{z}}+\dot{\bar{z}})^2+
\frac{\e ^{2s}}{z+\dot{z}}
(z\bar{z}+\bar{z}\dot{z}+\dot{z})^2+4\e ^{2s}z\bar{z},
\label{newz}
\end{equation}
where $I_{0}$ is an arbitrary real constant.
Numerical treatment of the problem with the aid of the new first
integral is currently under consideration \cite{numer}. Finally, the
following remark can be made: As the second-order complex differential
equation \eref{edoz} was shown to be reducible to a third-order real
one, there is some hope that the new first integral may result in
the reduction to a {\it second-order real} differential equation.     

\ack
Financial support by Direcci\'on General de Ense\~nanza Superior, Spain 
(Project PB95-0371) is gratefully acknowledged.

\end{document}